\documentclass[12pt]{article}
\usepackage{a4wide}
\usepackage{epsfig}
\usepackage{amsmath}

\def\gsim{\mathrel{\rlap{\raise 1.5pt \hbox{$>$}}\lower 3.5pt
\hbox{$\sim$}}}
\def\lsim{\mathrel{\rlap{\raise 2.5pt \hbox{$<$}}\lower 2.5pt
\hbox{$\sim$}}}

\newlength{\absize}
\begin{document}
  \thispagestyle{empty}
  \pagestyle{empty}
  \renewcommand{\thefootnote}{\fnsymbol{footnote}}
\newpage\normalsize
    \pagestyle{plain}
    \setlength{\baselineskip}{4ex}\par
    \setcounter{footnote}{0}
    \renewcommand{\thefootnote}{\arabic{footnote}}
\newcommand{\preprint}[1]{%
  \begin{flushright}
    \setlength{\baselineskip}{3ex} #1
  \end{flushright}}
\renewcommand{\title}[1]{%
  \begin{center}
    \LARGE #1
  \end{center}\par}
\renewcommand{\author}[1]{%
  \vspace{2ex}
  {\Large
   \begin{center}
     \setlength{\baselineskip}{3ex} #1 \par
   \end{center}}}
\renewcommand{\thanks}[1]{\footnote{#1}}
\begin{flushright}
IFT 2000-21 \\
{hep-ph/0101208} \\
January 2001
\end{flushright}
\vskip 0.5cm

\begin{center}
{\bf \Large Resolving SM-like scenarios via Higgs boson production
at a Photon Collider:\\ I. 2HDM {\it versus} SM}
%{\bf \Large Standard-Model-like scenarios in the 2HDM
%and Photon Collider potential}
\end{center}
\vspace{1cm}
\begin{center}
Ilya F. Ginzburg$^{a}$,
Maria Krawczyk$^{b}$ {\rm and}
Per Osland$^{c}$
\end{center}
%-----------------------------------
%   Address
%-----------------------------------
\vspace{1cm}
\begin{center}
$^a$ Sobolev Institute of Mathematics, SB RAS,
     Prosp.\ ac.\ Koptyug, 4, 630090 Novosibirsk, Russia\\
$^b$ Institute of Theoretical Physics, Warsaw University, Poland\\
$^c$ Department of Physics, University of Bergen,
     Allegt.\ 55, N-5007 Bergen, Norway
\end{center}
\vspace{1cm}
%%%%%%%%%%%%%%%%%%%%%%%%%%%%%%%%%%%%%%%%%%%%%%%%%%%%%%%%%%%%%%

\begin{abstract}
We consider the possibility that after operations at the LHC and
$e^+e^-$ Linear Colliders a Higgs boson will be discovered,
but no signal of New Physics will be found (Standard-Model-like 
scenario). This can occur in the Standard Model (SM) as well
as in other models, including the Two-Higgs-Doublet Model
(2HDM), the MSSM, etc. Experiments at a Photon Collider can 
resolve these cases. 

In this paper we compare the SM and the 2HDM~(II).
In the analysis we use as independent quantities the ratios, to
the SM values, of couplings of the observed Higgs boson to gauge
bosons, and to up and down-type quarks ({\em basic couplings}).
We derive a relation between these ratios within the 2HDM~(II),
a {\it pattern relation}. With the aid of this relation, different
possible realizations of an SM-like scenario are found.
For these realizations, we calculate the loop couplings of the
Higgs boson with $\gamma\gamma$ and $Z \gamma$,
and also with gluons, taking into account the expected accuracy
in the measurements of the basic couplings.
The obtained deviation of the
two-photon width from its SM value is generally higher than the
expected inaccuracy in the measurement of $\Gamma_{\gamma \gamma}$
at a Photon Collider.  The result is sensitive to the parameters of
the Higgs self interaction.
\end{abstract}

%%%%%%%%%%%%%%%%%%%%%%%%%%%%%%%%%%%%%%%%%%%%%%%%%%%%%%%%%%%%%%%%%%%%%
\section{Introduction}
%%%%%%%%%%%%%%%%%%%%%%%%%%%%%%%%%%%%%%%%%%%%%%%%%%%%%%%%%%%%%%%%%%%%%

In a widely discussed optimistic scenario, new physics will
reveal itself immediately above the Fermi scale $v=246$ GeV, and
new particles will be sufficiently light that they can be
discovered at the Tevatron and the LHC. Linear Colliders,
including a Photon Collider mode ($\gamma\gamma$ and $e\gamma$
\cite{GKST}), would be machines for measuring precise values of
coupling constants and exploring in detail supersymmetry, etc.\
\cite{Accomando}.

However, it could happen that the Higgs boson $h$ will be discovered,
but no signal of New Physics will be found, beyond the Standard Model
(SM). In this {\em SM-like scenario}, in particular, the Higgs
boson partial widths or coupling constants squared are precisely
measured, being in agreement with the SM within the experimental
accuracies. This can happen not only in the SM, but also if
Nature is described by some other theory, for example,
the Two-Higgs-Doublet Model (2HDM) or the
Minimal Supersymmetric Standard Model (MSSM). In this case the
main task for new colliders will be to search for signals of new
physics via deviations of observed quantities from SM
predictions. After LHC and $e^+e^-$ Linear Collider operations
the study of Higgs boson production at a Photon Collider
offers excellent opportunities for this \cite{sitges}.
Indeed, in the SM and in its extensions, all fundamental charged
particles contribute to the $h\gamma \gamma$ and $hZ\gamma$
effective couplings which can be tested at such a collider.
Besides, these couplings are absent in the
SM at tree level, appearing only at the loop level. Therefore,
the background for signals of new physics will be relatively
lower here than in processes which are allowed at tree level of
the SM. We focus our consideration on the era after LHC and
$e^+e^-$ Linear Collider operations, all numbers for
experimental uncertainties are related to this period. However, we
discuss additionally some effects for the preceding era (but after
discovery of the Higgs boson).

In the 2HDM and MSSM the observed Higgs boson can be
either one of two neutral CP-even scalars, $h$ or $H$
($M_h<M_H$).\footnote{Discussing these scalars, we use the
notation $\phi$ for both of them.}
In the present paper we compare the SM and the SM-like scenario
in the 2HDM in its Model~II variant of couplings with fermions,
2HDM~(II).
We derive a specific constraint
for the couplings of these scalars to fermions and gauge bosons,
valid in the 2HDM~(II) and MSSM --- {\em the pattern relation}.
It helps us to obtain additional, as compared to anticipated
future data,
constraints for the couplings of the observed Higgs scalar to
fermions and $W$ or $Z$ bosons.

To investigate how one can
resolve the SM and the 2HDM~(II) in the SM-like scenario, we
calculate the $\gamma\gamma$ and $Z\gamma$ partial widths of the
Higgs boson. We found that generally the $\gamma\gamma$ width
deviates sufficiently from the corresponding SM values that one
can distinguish models with the anticipated data, and measuring
the $Z\gamma$ width can support this conclusion.

Besides, for some sets of coupling constants which are possible
within the SM-like scenario, the models can be distinguished
via a study of the two-photon and (or) two-gluon widths even
before Photon Collider operations, at the Tevatron, the LHC or
an $e^+e^-$ Linear Collider.

In a forthcoming paper we plan to study the MSSM in this
same SM-like scenario, taking into account the anticipated
improvement in the precision of the Higgs boson two-photon width.

%%%%%%%%%%%%%%%%%%%%%%%%%%%%%%%%%%%%%%%%%%%%%%%%%%%%%%%%%%%%%%%%%%%%%
\section{Standard-Model-like scenario}\label{secSMlike}
%%%%%%%%%%%%%%%%%%%%%%%%%%%%%%%%%%%%%%%%%%%%%%%%%%%%%%%%%%%%%%%%%%%%%

We now consider the following scenario, referred to below as the
{\it SM-like scenario}. It is given by the following criteria
(whose precise formulations vary with time):\footnote{We discuss
below the era after LHC \cite{lhc} and $e^+e^-$ Linear Collider
operations.}
\begin{enumerate}
\item
One Higgs boson will be discovered (with mass above today's
limit for an SM Higgs boson, 115~GeV \cite{LEPC}). It can be
either the Higgs boson of the SM or one out of several neutral
scalars of some other model, such as the 2HDM or the MSSM.

\item  {\em No other Higgs boson will be discovered.} It means
that other Higgs bosons, if they exist, are either weakly 
coupled with the $Z$ and $W$ bosons, gluons and quarks, 
or sufficiently heavy to escape (direct or indirect) observation,
$M_H,\; M_A,\; M_{H^{\pm}}> {\cal O}(800\mbox{ GeV})$ \cite{lhc}.

\item All other new particles that may exist are heavier than
the discovery limits of LHC and the $e^+e^-$ Linear Collider.

\item The measured decay widths of this Higgs boson (or coupling
constants squared) to other particles, $\Gamma_i^{\rm exp}$,
will be in agreement with their SM values $\Gamma_i^{\rm SM}$
within the experimental precision $\delta_i$
\begin{equation}
\left|\frac{\Gamma_i^{\rm exp}}{\Gamma_i^{\rm SM}}-1\right|
\lsim\delta_i\ll 1.
\label{widthest}
\end{equation}
The values $\delta_i$ will change with time.  In our numerical
studies we use the most precise estimates for
the $e^+e^-$ Linear Collider, see Eq.~(\ref{couplacc}). 
Note that even if all $\Gamma^{\rm exp}_i/\Gamma^{\rm SM}_i$=1
for $i=q,\ell,W$ or $Z$, the SM could still be violated
(due to different signs of coupling constants and the existence
of additional particles like a charged Higgs boson).
\end{enumerate}

This scenario is realized, for example, for coupling constants
corresponding to the decoupling limit for a particular version of
the 2HDM~(II) \cite{Haber} or in the MSSM \cite{Haber,Djouadi1}.
The decoupling limit considered for the 2HDM~(II) or in the MSSM
is usually determined as that in which all coupling constants of
the lightest Higgs boson are very close to the SM values
irrespective of their experimental inaccuracies, assuming that
the observed neutral Higgs boson is the lightest one.
Besides, all other particles, Higgs particles and (in the MSSM)
super-particles, are assumed to be very heavy.
Further ``natural'' assumptions, which are valid in the MSSM but
not necessarily in the 2HDM, are added in the treatment of the
decoupling limit in Ref.~\cite{Haber}.
Such a situation can be realized in our SM-like scenario.
However, the SM-like scenario also permits
other realizations which for the 2HDM~(II) are discussed below.

%%%%%%%%%%%%%%%%%%%%%%%%%%%%%%%%%%%%%%%%%%%%%%%%%%%%%%%%%%%%%%%%%%%%%
\section{Couplings in the SM-like scenario}
%%%%%%%%%%%%%%%%%%%%%%%%%%%%%%%%%%%%%%%%%%%%%%%%%%%%%%%%%%%%%%%%%%%%%

%%%%%%%%%%%%%%%%%%%%%%%%%%%%%%%%%%%%%%%%%%%%%%%%%%%%%%%%%%%%%%%%%%%%%
\subsection{Expected precision of measured SM Higgs couplings}
%%%%%%%%%%%%%%%%%%%%%%%%%%%%%%%%%%%%%%%%%%%%%%%%%%%%%%%%%%%%%%%%%%%%%

Let $\delta_i$ be the relative experimental
uncertainty in the partial width of a considered Higgs boson,
$\Gamma_i^{\rm exp}$ (or coupling constants squared)
\begin{equation}
\label{Eq:delta}
\delta_i=\frac{\delta\Gamma_i^{\rm exp}}{\Gamma_i^{\rm exp}}.
\end{equation}
Actually, often combinations of partial widths are measured
(like branching ratios) and experimental estimates are obtained
for these quantities. Nevertheless, in our numerical study
(which should be updated after actual experimentation), we use
the values obtained for uncertainties of branching ratios etc.,
for the $\delta_i$ themselves.

At the LHC, the expected relative uncertainties (\ref{Eq:delta})
for the SM Higgs particle are of the order of 10--20\% \cite{lhc}.  
At the TESLA $e^+e^-$ collider the discussed production cross 
sections are expected to be measured with a significantly higher 
precision (at the same SM Higgs boson mass).  
At $M_h\le 140$~GeV (where the $b\bar{b}$ decay channel is 
dominant) and with integrated luminosity
500--1000~fb$^{-1}$ one can expect 
\cite{battaglia,Hagiwara:2000tk}:\footnote{For an up-dated
estimate for the low-mass region, see Ref.~\cite{Battaglia:2001jb}.}
\begin{equation}\begin{array}{llll}
\delta_b=0.024, \qquad &\delta_\tau=0.05,
\qquad  &\delta_c=0.083, \qquad &\delta_t=0.055,\\
\delta_Z=0.01, \qquad  &\delta_W=0.054,
\qquad  &\delta_g=0.055, \qquad &\delta_\gamma=0.19.
\end{array}\label{couplacc}
\end{equation}
In contrast to other quantities here, the numbers for $\delta_Z$
and $\delta_t$ are for coupling constants from the analysis of
the corresponding cross sections (not of branching ratios like
the others). The first one will be obtained via the study of
Higgs-strahlung by the recoil mass method, the second via
comparison of the cross section $e^+e^-\to t\bar{t}h$ with that
of Higgs-strahlung.  

Experiments at Photon Colliders open new perspectives. In
particular, even with an integrated luminosity of a
$\gamma\gamma$ collider in the high energy peak of about
$40$~fb$^{-1}$ (which is at least 5 times less than that
discussed in recent proposals \cite{hZga}),
a $\gamma\gamma$ collider makes
it possible to improve on the accuracy in measuring the
$h\gamma\gamma$ width  (via $b\bar{b}$ decay) up to
\cite{JikS}:
\begin{equation}
\delta_\gamma= 0.02\quad \mbox{ for }M_h<140\mbox{ GeV}\,.
\label{jik}
\end{equation}
This improvement is crucial for the considered problem.

At $M_h>140$~GeV the other decay channels and production 
mechanisms become important and the above uncertainties change 
strongly. The dependencies of some of these uncertainties on the 
Higgs boson mass are presented in Ref.~\cite{battaglia}.
For the Higgs boson production at a Photon Collider one should 
study it via the $WW^*$ and $ZZ^*$ decay modes (instead of 
$b \bar b$) to find the accuracy in the measurement of 
$\Gamma_{h\gamma\gamma}$.

The accuracy in the measurement of the effective $h Z\gamma$
coupling (in the process $e\gamma\to eh$) is evidently not so
high, it requires a separate study.

The value of the considered (SM-like) Higgs boson mass is
expected to be obtained with very high accuracy, 40--90~MeV
depending on the mass \cite{battaglia}.  Therefore, we will not
consider the uncertainty in the determination of the SM-width
arising from the inaccuracy in the Higgs boson mass.

%%%%%%%%%%%%%%%%%%%%%%%%%%%%%%%%%%%%%%%%%%%%%%%%%%%%%%%%%%%%%%%%%%%%%
\subsection{Limits on coupling constants from future data}
%%%%%%%%%%%%%%%%%%%%%%%%%%%%%%%%%%%%%%%%%%%%%%%%%%%%%%%%%%%%%%%%%%%%%

In our discussion we use quantities whose deviations from unity
for the observed Higgs boson provide some measure of whether the
Standard Model is realized or not. Such quantities are ratios of
actual (in principle measurable) coupling constants of each neutral
Higgs scalar $\phi$ ($h$ or $H$) with particle $i$ (or channel $i$)
to the corresponding value for the Higgs boson in the SM,
\begin{equation}
\label{Eq:chi-def}
\chi_i^\phi= \frac{g_i^\phi}{g_i^{\rm SM}}\,.
\end{equation}

In the SM-like scenario, for the observed Higgs boson 
the $|\chi_i|$ are close to 1, i.e.,
\begin{equation}
\chi_i^{\rm obs}=\pm (1-\epsilon_i),\quad
\mbox{with }  |\epsilon_i|\ll 1\,.
\label{estacc}
\end{equation}
The allowed ranges for $\epsilon_i$ are constrained by the
experimental accuracies $\delta_i$. In our SM-like scenario the
measured value of each relative width differs from unity by a
value no larger than $\delta_i$, with their central values
having uncertainties $\delta_i$. Therefore, the physical values
of the relative widths (\ref{Eq:chi-def}) can differ from 1 by
at most $2\delta_i$, and since $|\chi_i^2-1|\simeq 2|\epsilon_i|$, we have
\begin{equation}
|\epsilon_i|\le \delta_i\,.  \label{estacc1}
\end{equation}
Below we find additional constraints to these $\epsilon_i$ which
follow from the different realizations of
an SM-like scenario in the 2HDM~(II).

%%%%%%%%%%%%%%%%%%%%%%%%%%%%%%%%%%%%%%%%%%%%%%%%%%%%%%%%%%%%%%%%%%%%%
\section{Two-Higgs-Doublet Model (II)}
%%%%%%%%%%%%%%%%%%%%%%%%%%%%%%%%%%%%%%%%%%%%%%%%%%%%%%%%%%%%%%%%%%%%%

We limit our considerations to the CP-conserving Two-Higgs-Doublet
Model in its Model~II implementation, denoted by 2HDM~(II)
\cite{Hunter,barroso}.  Here, one doublet of fundamental scalar
fields couples to $u$-type quarks, the other to $d$-type quarks
(and charged leptons). The Higgs sector contains three neutral
Higgs particles, two CP-even scalars $h$ and $H$, one CP-odd
(pseudoscalar) $A$, and charged Higgs bosons $H^\pm$. We adopt
a scalar Higgs potential parameterized as in
Refs.~\cite{Hunter,Djouadi1}.\footnote{Another form of the scalar
potential is used in Ref.~\cite{Haber}.
It includes an additional constraint among the parameters and
a new degree of freedom governed by an additional parameter.
In this sense, the 2HDM of \cite{Haber} differs from that of
\cite{Hunter}.}

If the SM-like scenario is realized in the 2HDM we need to
consider both possibilities: not only the light scalar Higgs
boson, $h$, but also the heavier one, $H$, could imitate the SM
Higgs boson if the lighter scalar $h$ escapes detection (see
\cite{MK,GUN}). Therefore we discuss the coupling constants for
both neutral Higgs scalars.

%%%%%%%%%%%%%%%%%%%%%%%%%%%%%%%%%%%%%%%%%%%%%%%%%%%%%%%%%%%%%%%%%%%%%
\subsection{Basic couplings}
%%%%%%%%%%%%%%%%%%%%%%%%%%%%%%%%%%%%%%%%%%%%%%%%%%%%%%%%%%%%%%%%%%%%%

The ratios, relative to the SM values, of the direct coupling 
constants of the Higgs boson
$\phi=h$ or $H$ to the gauge bosons $V=W$ or $Z$, to up and
down quarks and to charged leptons {\it (basic couplings)}
can be determined via angles $\alpha$ and $\beta$ \cite{Hunter}:
\begin{equation}\begin{array}{ll}
\chi_V^h=\sin(\beta-\alpha),& \chi_V^H=\cos(\beta-\alpha),  \\
\chi_u^h=\sin(\beta-\alpha)+\cot\beta\cos(\beta-\alpha),\qquad
&\chi_u^H=\cos(\beta-\alpha)-\cot\beta\sin(\beta-\alpha), \\
\chi_d^h =\sin(\beta-\alpha)-\tan\beta\cos(\beta-\alpha),&
\chi_d^H =\cos(\beta-\alpha)+\tan\beta\sin(\beta-\alpha),  \\
\end{array}
\label{2hdmcoup-h}
\end{equation}
and for the CP-odd Higgs boson $A$
\begin{equation}
\chi_V^A=0,\qquad
\chi^A_u=-\cot\beta,\qquad
\chi^A_d=-\tan\beta. \nonumber
\end{equation}
(The normalization for $A$ is similar to that for $h$ or $H$, with an
additional $\gamma_5$ factor.) Here $\beta$ parameterizes the
ratio of the vacuum expectation values of the two Higgs
doublets and $\alpha$ parameterizes mixing among the two neutral
CP-even Higgs fields.  The angle $\beta$ is usually chosen in
the range $(0,\,\pi/2)$ and the angle $\alpha$ in the range
$(-\pi,\;0 )$.  (In much of the literature, the angle $\alpha$
is {\it incorrectly} taken in a narrower range, $(-\pi/2,\,0)$.)

The coupling of the charged Higgs boson to the neutral scalars
$\phi$ depends on the Higgs-boson masses and on the additional
parameter $\lambda_5$ \cite{Hunter,Djouadi1}.  We write this
coupling, in units of the coupling of the Higgs particle $\phi$
to arbitrary scalar particles with mass $M_{H^{\pm}}$ added to
the SM, $-2iM_{H^\pm}^2/v$ (compare \cite{Djouadi1}):
\begin{equation}
\chi_{H^\pm}^\phi
=\left(1-\frac{M_\phi^2}{2M_{H^\pm}^2}\right)\chi_V^{\phi}
+\left(\frac{M_\phi^2}{2M_{H^\pm}^2}-\frac{\lambda_5}{2\lambda_4}\right)
(\chi_u^{\phi}+\chi_d^{\phi}), \quad
\lambda_4=\frac{2M_{H^\pm}^2}{v^2},
\label{b2d2}
\end{equation}
where $v=(\sqrt{2}G_{\rm F})^{-1/2}=246$~GeV is the the vacuum
expectation value of the SM Higgs field. (Note that this
parameter $\lambda_5$ \cite{Hunter} differs from that considered
in \cite{Haber}. In particular, in the MSSM case
$\lambda_5=2M_A/v^2$ in \cite{Hunter} while $\lambda_5=0$ in
\cite{Haber}.)

%%%%%%%%%%%%%%%%%%%%%%%%%%%%%%%%%%%%%%%%%%%%%%%%%%%%%%%%%%%%%%%%%%%%%
\subsection{Pattern relation}
%%%%%%%%%%%%%%%%%%%%%%%%%%%%%%%%%%%%%%%%%%%%%%%%%%%%%%%%%%%%%%%%%%%%%

The ratios $\chi_i^\phi$ of Eq.~(\ref{2hdmcoup-h}), for the basic
couplings of each scalar, are more closely related to the
observables and in the forthcoming analysis we use them, instead
of the parameters $\alpha$ and $\beta$.  Since for each $\phi$ these
three $\chi_i$ can be expressed in terms of {\it two} angles, they
fulfill a simple relation {\em(pattern relation)}, which plays a
basic role in our analysis.\footnote{A similar idea was explored
in \cite{GUN} to construct {\em sum rules} for quantities like
$\chi_i^2$, relating the production cross sections of the Higgs
boson at an $e^+e^-$ collider in different channels.} It has
the same form for both $h$ and $H$, namely $(\chi_u -\chi_V)
(\chi_V -\chi_d) +\chi_V^2=1$, or
\begin{equation}
(\chi_u +\chi_d)\chi_V=1+\chi_u \chi_d.
\label{2hdmrel}
\end{equation}
It is useful to note additionally that from
Eq.~(\ref{2hdmcoup-h}) follows an expression for $\tan\beta$:
\begin{equation}
\label{Eq:tan-beta}
\tan^2\beta={\frac{\chi_V-\chi_d}{\chi_u-\chi_V}}
={\frac{1-\chi_d^2}{\chi_u^2-1}}.
\end{equation}
(Note that $ 1>\chi_d^2$ and $\chi^2_u>1$, or vice versa.)

The pattern relation (\ref{2hdmrel}) was obtained here in the
tree approximation. Radiative corrections will modify it.  We
discuss their effect at the one-loop level.  The largest gluon
corrections to the $\phi q\bar{q}$ vertices are identical in the SM
and in the 2HDM. They cancel in the ratios $\chi_u$ and
$\chi_d$. The electroweak corrections to $\chi_i$ include, in
particular, contributions with Higgs bosons in the loops which
depend on the Higgs self-interaction constants.  Let us recall
that after fixing the Higgs boson masses, one free parameter,
$\lambda_5$, remains in the 2HDM~(II).  It results in a
theoretical uncertainty $\propto \lambda_5$ in the discussed
electroweak corrections.  Both these corrections themselves and
the uncertainty in them are of the order of $\alpha_{\rm em}$.
This makes it impossible to fix the pattern relations with an
accuracy better than a few per cent even at moderate values of
$\lambda_5$ without a detailed analysis of possible
cancellations among various contributions.

%%%%%%%%%%%%%%%%%%%%%%%%%%%%%%%%%%%%%%%%%%%%%%%%%%%%%%%%%%%%%%%%%%%%%
\subsection{Perturbativity constraints}
%%%%%%%%%%%%%%%%%%%%%%%%%%%%%%%%%%%%%%%%%%%%%%%%%%%%%%%%%%%%%%%%%%%%%

All estimates below are within the lowest order of
perturbation theory (tree level or one loop), assuming in fact
that the loop corrections are small. Therefore, coupling
constants of the theory (at least, those related to the discussed
phenomena) should not be too large (``perturbativity'' condition
\cite{pert}). Since the effective parameters of perturbation
theory are $g^2/(4\pi)^2$ for Yukawa couplings and $\lambda/(4\pi)^2$
for the scalar self-interaction,
these perturbativity conditions are generally not very restrictive,
$g<{\cal O}(4\pi)$ and $\lambda<{\cal O}(16\pi^2)$.
In the SM, for example, the conditions correspond to an upper
mass value of fourth-generation quarks of about 3~TeV.

In the 2HDM~(II) and within our SM-like scenario, these
constraints in general give limitations for the couplings of
the {\it observed} and {\it unobserved} Higgs bosons.
If the observed Higgs boson is the lighter one, $h$, the possible
strong coupling of the other Higgs boson $H$ to quarks is irrelevant
to our problem since it is too heavy to be observable. On
the other hand, if the observed Higgs boson is the heavier one, $H$,
then the possible strong coupling of the light Higgs boson $h$
should be considered since it is in principle within
the domain of observation. When perturbativity is violated for this
light, unobserved Higgs boson, the estimates of its couplings according to
Eq.~(\ref{2hdmcoup-h}) become incorrect. In this case, for
example, the limitation $g_{t\bar{t}\phi}<{\cal O}(4\pi)$ gives
$\tan\beta>{\cal O}(m_t/4\pi v)\approx 0.05$ and the limitation
$g_{b\bar{b}\phi}<{\cal O}(4\pi)$ gives 
$\tan\beta<{\cal O}(4\pi v/m_b)\approx 600$. 
With effective parameters of the perturbation series lower than 
0.1, these estimates give $\tan\beta>0.15$ -- 0.2 for the
$t\bar{t}\phi$ coupling and $\tan\beta<150$ -- 200 for the
$b\bar{b}\phi$ coupling (see Ref.~\cite{pert} for details).
With a scalar potential in the form of \cite{Hunter} the
perturbativity conditions on the Higgs masses are $M_A$, $M_H$,
$M_{H^\pm}\lsim 3$~TeV. Note that, in fact, in the 2HDM these estimates
are oversimplified. Indeed, the scalar potential here has
a complex form, some coefficients can have opposite signs.
In loop corrections different couplings can partly compensate each
other, making perturbativity limitations less restrictive,
and some of the scalars $A$, $H$,
$H^\pm$ can even be heavier than 3 TeV without violating
perturbativity.

%%%%%%%%%%%%%%%%%%%%%%%%%%%%%%%%%%%%%%%%%%%%%%%%%%%%%%%%%%%%%%%%%%%%%
\subsection{2HDM~(II) vs.\ MSSM}
%%%%%%%%%%%%%%%%%%%%%%%%%%%%%%%%%%%%%%%%%%%%%%%%%%%%%%%%%%%%%%%%%%%%%

The Higgs sector in the MSSM represents a specific realization of
that in the 2HDM~(II). Here we only point out the common points as well
as differences between the general 2HDM~(II) and the MSSM, which are 
essential for our discussion.

The Higgs sector and couplings of Higgs bosons to quarks and gauge
bosons expressed in terms of mixing angles (\ref{2hdmcoup-h}) and
the relation between them (like the pattern relation, Sec.~4.2)
are identical for the 2HDM~(II) and the MSSM.
However, in the MSSM there are additional constraints.
In the MSSM the lightest Higgs boson mass is restricted by
$M_h\lsim135$ GeV \cite{MSSM}. 
For $M_H\gg M_h$ one has $M_A\approx M_H\approx M_{H^\pm}$,
while in the 2HDM the masses of the Higgs bosons and the angles 
$\alpha$ and $\beta$ are free parameters. 
In the MSSM the $\alpha$ range is ($-\pi/2,\,0$), 
to be compared to ($-\pi,\,0$) in the 2HDM, and
$\sin(\beta-\alpha)>0$. In the MSSM the sign of
$\cos(\beta-\alpha)$ coincides with that of $1-\tan\beta$ while
in the 2HDM both signs of $\cos(\beta-\alpha)$ are allowed.
Finally, in the MSSM $\lambda_5=2M^2_{A}/v^2$, and therefore, 
if $A$ is heavy, $\lambda_5\approx \lambda_4$, see Eq.~(\ref{b2d2}).

%%%%%%%%%%%%%%%%%%%%%%%%%%%%%%%%%%%%%%%%%%%%%%%%%%%%%%%%%%%%%%%%%%%%%%%%
\section{SM-like realizations in the 2HDM~(II)}
%%%%%%%%%%%%%%%%%%%%%%%%%%%%%%%%%%%%%%%%%%%%%%%%%%%%%%%%%%%%%%%%%%%%%%%%

\subsection{General discussion}

The SM-like scenario is determined by several natural criteria
(Sec.~2). 
We start with a discussion of regions in the parameter
space of the 2HDM~(II) where this scenario can be realized, 
considering the criterion on SM-like couplings (criterion 4). 
Next we consider the possible observation of a heavier Higgs
boson in the associated production with $t$ or $b$ quarks.
Simultaneously we consider whether other Higgs bosons necessarily
are very heavy (criterion 2 --- ``no discovery'' of other Higgs 
bosons). Besides, it can happen that the observation of 
the loop-induced two-gluon or two-photon Higgs decay at a hadron
or an $e^+e^-$ collider will show that the SM-like scenario is
violated even though the basic couplings or basic widths
are close to the SM values. To study this possibility we
calculate in Sec.~\ref{secloop} the loop-induced couplings for
all considered solutions and discuss the corresponding
distinguishability of models in Sec.~\ref{secdist}.

Assuming that the basic couplings of the considered Higgs boson
(with $i=u,d,V$) satisfy the SM-like scenario, we solve
Eqs.~(\ref{estacc}) in terms of $\epsilon_V, \epsilon_d$ and
$\epsilon_u$, under the constraints of the pattern relation
(\ref{2hdmrel}). We found two basic classes of solutions denoted
$A_{\phi}$ and $B_{\phi}$.

First, for the observed Higgs boson $\phi$ there are solutions
with approximately identical couplings $\chi_V\approx \chi_u\approx
\chi_d\approx \pm 1$, denoted $A_{\phi\pm}$. Here, the first
subscript labels the observed Higgs boson ($\phi=h,H$).
The second subscript $\pm$ labels the sign of $\chi_V^\phi$.
For example, the solution $A_{H-}$ is that with
the observed Higgs boson being the heavier one, $H$, and with
$\chi_V^H\approx -1$. The solutions $A$ are really close to
the SM for all basic couplings: relative phases coincide, and
magnitudes are practically the same.  There are also solutions
denoted $B_{\phi \pm}$, where some of the $\chi_i\approx 1$ but
other $\chi_j\approx -1$.  The subscripts here have the same
meaning as above.  These solutions $B$ are in fact distinct from
the SM case, even though all basic widths $\sim|\chi_i|^2$ are
close to the SM values.  Note that, by definitions
(\ref{estacc}) and (\ref{2hdmcoup-h}), in all cases
$\epsilon_V>0$ while the signs of $\epsilon_u$ and $\epsilon_d$
cannot be fixed in advance.

%%%%%%%%%%%%%%%%%%%%%%%%%%%%%%%%%%%%%%%%%%%%%%%%%%%%%%%%%%%%%%%%%%%%%%%%
\subsubsection{\boldmath$h$ as an SM-like Higgs boson}
%%%%%%%%%%%%%%%%%%%%%%%%%%%%%%%%%%%%%%%%%%%%%%%%%%%%%%%%%%%%%%%%%%%%%%%%

For the solutions $A_{h\pm}$ and $B_{h\pm}$, the observed Higgs
boson is $h$, with $\sin(\beta-\alpha)\approx \pm 1$, so we have
$\cos (\beta-\alpha)\approx 0$. For these solutions the $H$
boson can escape observation even if it is not very heavy.
Indeed, in these cases the coupling $HVV$, proportional to
$\cos(\beta-\alpha)$, Eq.~(\ref{2hdmcoup-h}), can be so small
that $H$ is neither observed in Higgs-strahlung nor in
$WW$-fusion at an $e^+e^-$ Linear Collider. However, for
$\tan\beta<1$ the $Ht\bar{t}$ coupling can be significant so
that the non-discovery of this Higgs boson in $t\bar{t}H$
production (and gluon fusion)
at the LHC or an $e^+e^-$ Linear Collider allow
to obtain a constraint for $M_H$ (from below) within
an SM-like scenario. The same should hold for the case
$\tan\beta\gg 1$ and $b\bar{b}H$ production at the LHC (via
gluon fusion) or an $e^+e^-$ Linear Collider.

Note that very small values of $\tan\beta$ which are ruled out by
the perturbativity constraint of the $t\bar{t}H$ coupling are
irrelevant if $H$ is very heavy and escapes detection.

%%%%%%%%%%%%%%%%%%%%%%%%%%%%%%%%%%%%%%%%%%%%%%%%%%%%%%%%%%%%%%%%%%%%%%%%
\subsubsection{\boldmath$H$ as an SM-like Higgs boson}
%%%%%%%%%%%%%%%%%%%%%%%%%%%%%%%%%%%%%%%%%%%%%%%%%%%%%%%%%%%%%%%%%%%%%%%%

For the solutions $A_{H\pm}$ and $B_{H\pm}$ the observed Higgs
boson is $H$. For this case one should consider also the
coupling constants of the light Higgs boson $h$ in more detail to
check a basic point of the SM-like scenario: criterion~2 of
Sec.~2 --- {\em no other Higgs particle should be discovered}.
Let us discuss the corresponding constraints.

For these solutions we have $\cos(\beta-\alpha)\approx \pm 1$
and $\sin(\beta-\alpha)\approx 0$. Therefore, the coupling of
the lightest Higgs boson $h$ to gauge bosons, proportional to
$\sin (\beta-\alpha)$, Eq.~(\ref{2hdmcoup-h}), can be so small
that $h$ is neither observed in Higgs-strahlung nor in
$WW$-fusion at an $e^+e^-$ Linear Collider.

If $\tan\beta<1$, the coupling of the lightest Higgs boson $h$
with $t$-quarks becomes very large, and this Higgs boson cannot
avoid observation in the associated production in the
$t\bar{t}h$ channel, which would contradict the SM-like
scenario. Therefore, this case is ruled out from the family of 
SM-like realizations.
(With this restriction we also stay within the perturbative
region for the $Ht\bar{t}$ coupling.)

The case $\tan\beta\gg 1$ with strong $hb\bar{b}$ coupling can be
difficult for observation of the lightest Higgs boson in the
$hb\bar{b}$ channel due to high $b\bar{b}b\bar{b}$ background.
The extremely large values of $\tan\beta$ are suppressed in the
SM-like scenario since in this case the $h\gamma\gamma$ and
$hZ\gamma$ (loop) couplings can become so high that the lightest
Higgs boson can be observed in processes like $e^+e^-\to h\gamma$
at GigaZ \cite{MK}.  An analysis of the data for this
case shows that the current non-discovery window corresponds to
2~GeV $<M_h<20$~GeV, if $\tan\beta$ is below 20, with
$\sin^2(\beta-\alpha) <0.02$. For higher $M_h$ the window in
$\tan\beta$ is even higher (for $M_h$= 50~GeV, the maximal
allowed $\tan \beta=100$ but then $\sin^2(\beta-\alpha) <0.1$
\cite{MK,gun00}).

%%%%%%%%%%%%%%%%%%%%%%%%%%%%%%%%%%%%%%%%%%%%%%%%%%%%%%%%%%%%%%%%%%%%%%%%
\subsection{Solutions A -- allowed ranges for couplings}
\label{solAdescr}
%%%%%%%%%%%%%%%%%%%%%%%%%%%%%%%%%%%%%%%%%%%%%%%%%%%%%%%%%%%%%%%%%%%%%%%%

We first consider the solutions near the SM point for all basic
coupling constants, i.e.,
\begin{equation}
\chi_V=\pm(1-\epsilon_V)\;\approx\; \chi_d=\pm(1-\epsilon_d)\;
\approx\; \chi_u=\pm(1-\epsilon_u).
\label{AH}
\end{equation}

Using the pattern relation (\ref{2hdmrel}) and
Eq.~(\ref{Eq:tan-beta}), we obtain for all these solutions,
neglecting small terms of higher order in $\epsilon_i$,
\begin{equation}
\epsilon_u=-\frac{2\epsilon_V}{\epsilon_d}
\Rightarrow \epsilon_V=-\frac{\epsilon_u\epsilon_d}{2}
<\frac{\delta_u\delta_d}{2}, \quad
\tan\beta
=\sqrt{\left|\frac{\epsilon_d}{\epsilon_u}\right|}.
\label{sol2s}
\end{equation}

Since $\epsilon_V>0$, in all these solutions the signs of
$\epsilon_u$ and $\epsilon_d$ are opposite. Besides,
$\epsilon_V$ here is given by the product of two other
$\epsilon$'s, so that it should be extremely small,
($\epsilon_V\le0.001$ using $\delta_t$, $\delta_b$ from
Eq.~(\ref{couplacc})). Therefore, one can neglect this deviation
from 1, and in the calculations of loop-induced couplings and
coupling with the charged Higgs boson one can put $\chi_V=\pm
1$, irrespective of the experimental uncertainty $\delta_V$
($\delta_Z$).
Finally, it is useful to note that in the
cases when $\tan\beta>1$ one should have $|\epsilon_d|>|\epsilon_u|$, 
regardless of the experimental uncertainties $\delta_{u,d}$. 
It gives additional constraint for the quantity $\chi_u$ 
which is usually measured with lower precision than $\chi_d$. 
In the case when $\tan\beta<1$ the opposite relation
$|\epsilon_d|> |\epsilon_u|$ is enforced independent of
experimental uncertainties, it can be useful at $M_h>140$ GeV.

%%%%%%%%%%%%%%%%%%%%%%%%%%%%%%%%%%%%%%%%%%%%%%%%%%%%%%%%%%%%%%%%%%%%%%%%
\subsubsection{\boldmath Solutions $A_{h\pm}$}
%%%%%%%%%%%%%%%%%%%%%%%%%%%%%%%%%%%%%%%%%%%%%%%%%%%%%%%%%%%%%%%%%%%%%%%%

\begin{itemize}
\item[$A_{h+}$:]
For this solution, $\sin(\beta-\alpha)\approx 1$, hence
$\beta-\alpha\approx \pi/2$.  In this case $\cos(\beta-\alpha)$
can be either negative with $\epsilon_d<0$ or positive with
$\epsilon_d>0$.

\item[$A_{h-}$:] Here, we have $\beta-\alpha\approx 3\pi/2$ with
$\beta-\alpha<3\pi/2$. (Therefore, $\cos(\beta-\alpha)$ is small
and negative in this case.) Within the allowed ranges of mixing
angles discussed in Sec.~4, solutions are at $\alpha \approx -\pi$,
$\beta\approx \pi/2$, and, therefore, $\tan\beta\gg 1$.
\end{itemize}

%%%%%%%%%%%%%%%%%%%%%%%%%%%%%%%%%%%%%%%%%%%%%%%%%%%%%%%%%%%%%%%%%%%%%%%%
\subsubsection{\boldmath Solutions $A_{H\pm}$}
%%%%%%%%%%%%%%%%%%%%%%%%%%%%%%%%%%%%%%%%%%%%%%%%%%%%%%%%%%%%%%%%%%%%%%%%

\begin{itemize}
\item[$A_{H+}$:]
This solution corresponds to $\cos(\beta-\alpha) \simeq 1$, thus
$\beta-\alpha\approx 0$.  Since $\beta>0$ and $\alpha <0$, this
case can occur only if $\alpha\approx 0$ and $\beta\approx 0$,
or $\tan\beta\ll 1$. The last inequality contradicts the
SM-like scenario as was noted above.
\item[$A_{H-}$:] This solution allows a positive
$\sin(\beta-\alpha)$ with negative $\epsilon_u$, or vice versa.
\end{itemize}

The allowed solutions $A_{\phi\pm}$ are summarized in 
Table~\ref{tab1}. 
%================================================================
\begin{table}[htb]
\begin{center}
\begin{minipage}{160mm}
\begin{tabular}{||c|c|c|c|c|c|c||}
\hline\hline & $\chi_V$ & $\chi_d$ & $\chi_u$ 
&$\sin(\beta-\alpha)$ & $\cos(\beta-\alpha)$  & $\tan\beta$
\\\hline\hline
$A_{h+}$ & $+(1-\epsilon_V)$ & $+(1-\epsilon_d)$ & $+(1-\epsilon_u)$ 
&$\approx1$ & \parbox[c]{25mm}{$\lsim 0\Rightarrow$ $\epsilon_d<0$
\\$\gsim 0\Rightarrow\epsilon_d>0$} &  $<1$ and $>1$\\ \hline
$A_{h-}$ & $-(1-\epsilon_V)$ & $-(1-\epsilon_d)$ &$-(1-\epsilon_u)$ 
& $\approx-1$ & $\lsim 0$ & $\gg 1$\\ \hline
$A_{H-}$& $-(1-\epsilon_V)$ & $-(1-\epsilon_d)$ &$-(1-\epsilon_u)$ 
&\parbox[c]{25mm}{ $\gsim 0\Rightarrow\epsilon_u<0$ \\  
$\lsim 0\Rightarrow\epsilon_u>0$} &$\approx-1$& $>1$  \\
\hline\hline
\end{tabular} \end{minipage} \end{center}
\caption{SM-like scenarios $A$ in the 2HDM~(II).
For each realization, $\epsilon_u$ and $\epsilon_d$ have opposite signs, 
and $\epsilon_V=-\epsilon_d\epsilon_u/2>0$ is negligible.}
\label{tab1}
\end{table}
%================================================================

%%%%%%%%%%%%%%%%%%%%%%%%%%%%%%%%%%%%%%%%%%%%%%%%%%%%%%%%%%%%%%%%%%%%%%%%
\subsection{Solutions B -- allowed ranges for couplings}\label{solBdescr}
%%%%%%%%%%%%%%%%%%%%%%%%%%%%%%%%%%%%%%%%%%%%%%%%%%%%%%%%%%%%%%%%%%%%%%%%

Solutions $B$ have different signs of
the basic couplings as compared to those of the SM.
We label these solutions by an additional subscript 
denoting the type of quark whose coupling with the observed 
Higgs boson is of opposite sign as compared with the gauge boson 
coupling, $\chi_V$.  We consider therefore solutions 
$B_{\phi\pm d}$ and $B_{\phi\pm u}$.  The solutions with
$-\chi_V\approx \chi_d\approx \chi_u\approx \pm 1$ cannot be
realized by virtue of Eq.~(\ref{2hdmcoup-h}).

According to Eq.~(\ref{b2d2}), for these solutions the coupling of
the charged Higgs boson $H^\pm$ to the observed neutral one,
$\chi_{H^\pm}$, is practically independent of $\lambda_5$. Also,
if the charged Higgs boson $H^\pm$ is heavy (as is the case
in the SM-like scenario), its coupling to the neutral
Higgs scalars $\phi$ is close to that of the vector bosons,
$\chi_{H^\pm} \approx \chi_V$.

%%%%%%%%%%%%%%%%%%%%%%%%%%%%%%%%%%%%%%%%%%%%%%%%%%%%%%%%%%%%%%%%%%%%%%%%
\subsubsection{\boldmath Solutions $B_{\phi\pm d}$}
%%%%%%%%%%%%%%%%%%%%%%%%%%%%%%%%%%%%%%%%%%%%%%%%%%%%%%%%%%%%%%%%%%%%%%%%

We have
\begin{equation}
\chi_V=\pm(1-\epsilon_V)\;\approx\;-
\chi_d=\pm(1-\epsilon_d)\;\approx\; \chi_u=\pm(1-\epsilon_u).
\label{BHd}
\end{equation}
Neglecting terms of higher order in $\epsilon$'s, we get
\begin{equation}
\epsilon_u=-\frac{\epsilon_V\epsilon_d}{2}\Rightarrow
|\epsilon_u|<\frac{\delta_V\delta_d}{2}, \qquad
\tan\beta\approx\sqrt{\frac{2}{\epsilon_V}}.
\label{solBds}
\end{equation}
The last equations show that this solution requires {\it large}
values of $\tan\beta$.

According to Eq.~(\ref{solBds}) and limits from
Eq.~(\ref{couplacc}) for $V=Z$ and $d=b$, the quantity
$\epsilon_u$, which in these solutions is given by the product of 
$\epsilon_V$ and $\epsilon_d$, should be extremely small.
Therefore, in the calculations of the loop-induced couplings,
one can put $\epsilon_u=0$ irrespective of the experimental 
uncertainty $\delta_u$ (e.g.\ $\delta_t$).

\begin{itemize}
\item[$B_{h+d}$:]
For this solution the observed Higgs boson is $h$ and
$\sin(\beta-\alpha) \approx 1$. With $\tan\beta>1$ this solution
can be realized if also $\cos(\beta-\alpha)>0$.
\item[$B_{h-d}$:] For this solution the observed Higgs boson is $h$
and $\chi_V =\sin(\beta-\alpha)\approx -1$ with
$\beta \lsim \pi/2$ and $\alpha \gsim -\pi$. Therefore, 
$\beta-\alpha \lsim 3\pi/2$ and $\cos (\beta-\alpha)<0$.
\item[$B_{H+d}$:] For this solution the observed Higgs boson is $H$
and $\chi_V=\cos(\beta-\alpha)\approx 1$. Since $\alpha<0$ and
$\beta>0$, it can only be realized if $\alpha\approx 0$ and 
$\beta\approx 0$. The last condition contradicts the constraint
$\tan\beta>1$ required for this type of solution (i.e., for
the $h$ to escape observation).  Therefore,
this solution cannot be realized in the SM-like scenario.  
\item[$B_{H-d}$:] For this solution the observed Higgs boson is 
$H$ and $\chi_V=\cos(\beta-\alpha)\approx -1$. This type of
solution can be realized with $\sin(\beta-\alpha)>0$.
\end{itemize}

%%%%%%%%%%%%%%%%%%%%%%%%%%%%%%%%%%%%%%%%%%%%%%%%%%%%%%%%%%%%%%%%%%%%%%%%
\subsubsection{\boldmath Solutions $B_{\phi\pm u}$}
%%%%%%%%%%%%%%%%%%%%%%%%%%%%%%%%%%%%%%%%%%%%%%%%%%%%%%%%%%%%%%%%%%%%%%%%

We have
\begin{equation}
\chi_V=\pm(1-\epsilon_V)\;\approx\;
\chi_d=\pm(1-\epsilon_d)\;\approx\;-\chi_u=\pm(1-\epsilon_u).
\label{BHu}
\end{equation}
Neglecting terms of higher order in $\epsilon$'s, we get
\begin{equation}
\epsilon_d=-\frac{\epsilon_V\epsilon_u}{2}\Rightarrow
|\epsilon_d|<\frac{\delta_V\delta_u}{2}\;, \qquad
\tan\beta\approx\sqrt{\frac{\epsilon_V}{2}}\,.
\label{solBus}
\end{equation}
The last equations show that this solution requires {\it small}
values of $\tan\beta$.

In this solution the quantity $\epsilon_d$ given by the product
of $\epsilon_V$ and $\epsilon_u$, should be extremely small
($|\epsilon_d|<0.0003$ according to the limits from 
Eq.~(\ref{couplacc}) for $V=Z$ and $u=t$) regardless of the 
$\delta_b$ value.

As was noted above, for solutions with the observed heavier Higgs 
boson, $B_{H\pm u}$, a small value of $\tan\beta$ contradicts the 
SM-like scenario since $h$ cannot escape observation.  The solution
$B_{h-u}$ cannot be realized either, since for this case we have
$\sin(\beta-\alpha) \approx -1$, implying $\alpha\approx-\pi$
and $\beta\approx \pi/2$. The last point contradicts the
required small value of $\tan\beta$, Eq.~(\ref{solBus}).
Therefore, in this case only one type of solution can be
realized, namely:
\begin{itemize}
\item[$B_{h+u}$:] For this solution the observed Higgs boson is 
$h$ and we have $\sin(\beta-\alpha) \approx 1$. With the 
condition $\tan\beta\ll 1$ this solution can be realized if also 
$\cos(\beta-\alpha)<0$.
\end{itemize}

The allowed solutions $B$ are summarized in Table~\ref{tab2}.
%================================================================
\begin{table}[ht]
\begin{center}
\begin{minipage}{140mm}
\begin{tabular}{||c|c|c|c|c|c|c||}
\hline\hline & $\chi_V$ & $\chi_d$ & $\chi_u$ &
$\sin(\beta-\alpha)$ & $\cos(\beta-\alpha)$  & $\tan\beta$
\\\hline\hline
$B_{h+d}$ & $+(1-\epsilon_V)$ & $-(1-\epsilon_d)$ & $+(1-\epsilon_u)$ 
&$\approx1$ & $\gsim 0$ &$>1$\\ \hline
$B_{h-d}$ & $-(1-\epsilon_V)$ &  $+(1-\epsilon_d)$& $-(1-\epsilon_u)$
&$\approx-1$ & $\lsim 0$ & $>1$\\ \hline
$B_{H-d}$ & $-(1-\epsilon_V)$ &$+(1-\epsilon_d)$ & $-(1-\epsilon_u)$
& $\gsim 0$ & $\approx-1$ & $>1$ \\ \hline\hline
$B_{h+u}$ & $+(1-\epsilon_V)$ & $+(1-\epsilon_d)$ & $-(1-\epsilon_u)$
&$\approx1$ & $\lsim 0$ &$\ll 1$ \\
\hline\hline
\end{tabular} \end{minipage} \end{center}
\caption{SM-like scenarios $B$ in the 2HDM~(II).
The solutions correspond to $\epsilon_V>0$ and, for each realization, 
to negative and positive $\epsilon_u$ and $\epsilon_d$ regions,
with $\epsilon_d\epsilon_u<0$.
For the solutions $B_{\phi\pm d}$, 
$\epsilon_u=-\epsilon_V\epsilon_d/2$ is negligible, whereas for
the solution $B_{h+u}$ $\epsilon_d=-\epsilon_V\epsilon_u/2$ 
is negligible.}
\label{tab2}
\end{table}
%================================================================

%%%%%%%%%%%%%%%%%%%%%%%%%%%%%%%%%%%%%%%%%%%%%%%%%%%%%%%%%%%%%%%%%%%%%%%%
\section{Loop-induced couplings in the SM-like scenario}\label{secloop}
%%%%%%%%%%%%%%%%%%%%%%%%%%%%%%%%%%%%%%%%%%%%%%%%%%%%%%%%%%%%%%%%%%%%%%%%

In this section we calculate the ratios $\chi_a$ of loop-induced 
couplings of the Higgs boson to photons and gluons in the 2HDM~(II) 
and the SM within the constraints obtained in the previous 
section and within the ranges of the basic coupling constants 
allowed by the expected experimental inaccuracies $\delta_i$ 
given for the post $e^+e^-$ Linear Collider era by 
Eq.~(\ref{couplacc}).

Up to model-independent factors, these couplings are proportional
to the quantities $F^a$ with $a=\gamma\gamma$, $Z\gamma$ or $gg$.
In the one-loop approximation $F^a$ is a sum of contributions
$F^a_J(P)$ from loops given by particles $P$ (the subscript $J$ 
labels its spin, see, e.g., in \cite{Hunter}).
To clarify the obtained result, we will discuss explicitly
the cases $a=\gamma\gamma$ and $a=gg$.
For these cases, the functions $F^a_J(P)$ are 
either identical (for $J=1/2$) or apply only for $a=\gamma\gamma$,
so we omit superscript $a$ for them. (The corresponding
functions for $a=Z\gamma$ with additional dependence on $Z$ 
virtuality necessary for calculation of $e\gamma\to e\phi$ 
cross section can be found, for example, in \cite{GIv1}.)
In the 2HDM we have
\begin{equation}
F^{\gamma\gamma} =\chi_W F_1(W) +
\sum\limits_{f=q,\ell} N_c Q_f^2 \chi_fF_{1/2}(f) +
\chi_{H^\pm}F_0(H^\pm), \label{phig}
\end{equation}
where $N_c=3$ for quarks, and 1 for leptons,
and a similar expression for $F^{Z\gamma}$.
Furthermore,
\begin{equation}
F^{gg}=\sum\limits_{q} N_c \chi_q F_{1/2}(q).
\label{ghig}
\end{equation}
The (in general complex) functions
$F_J(P)$  depend on the ratio $z_P=4M_P^2/M_\phi^2$ only. For the
considered region of Higgs-boson mass, only the contributions of
the $W$ boson, $t$ and $b$-quark are essential. Moreover, the
$b$-quark-loop contribution is very small as compared to the $W$-loop 
contribution. Therefore, for estimates 
one can write with high accuracy
\begin{equation}
\chi_{\gamma\gamma}=\frac{\chi_V
F_1(W) +\frac{4}{3}\chi_u F_{1/2}(t)
+\chi_{H^\pm}F_{0}(H^\pm)}{F_1(W) +\frac{4}{3}
F_{1/2}(t)}\,,\quad
\chi_{gg}=\frac{\chi_uF_{1/2}(t) +\chi_dF_{1/2}(b)}
{F_{1/2}(t)+F_{1/2}(b)}\,.
\label{loop}
\end{equation}
In the numerical calculations we use the complete equations
(\ref{phig}) and (\ref{ghig}).

For our discussion it is useful to present asymptotic values of
the loop integrals:
\begin{equation}\begin{array}{c}
F_1\to 7,\;\; F_{1/2}\to -\frac{4}{3},\;\;
F_0\to -\frac{1}{3}\;
\mbox{ for }z_P\gg 1,  \\
\\
F_{1/2}\to\frac{z}{2}\ln^2\frac{4}{z}\;
\mbox{ for }z_P\ll 1.
\end{array}
\label{asym}
\end{equation}
Note that $F_1(W)$ increases with $M_\phi$ to the $WW$
threshold where $F_1= 5+3\pi^2/4\approx 12.4$.

We present results of numerical calculations for the relative
widths $|\chi_a|^2$ ($a=\gamma\gamma$, $Z\gamma$ or $gg$) 
for solutions $A$ and $B$.
In the figures, solid curves correspond to the ``exact'' cases, 
where $|\chi_V|=|\chi_d| = |\chi_u|=1$.
The shaded bands in the figures around the solid curves are derived 
from the anticipated 1~$\sigma$ bounds for the measured
basic coupling constants, $g_V$, $g_u$ and $g_d$, with additional 
constraints given by the pattern relation for each solution 
as was discussed in Sec.~5. To obtain these
shaded regions, we varied each basic coupling entering these
widths, within the most narrow interval given by Eqs.~(\ref{couplacc})
and (\ref{estacc1}) for the coupling of a given type, namely
$\epsilon_V\le \delta_Z$, $|\epsilon_d|\le \delta_b$ and 
$|\epsilon_u|\le \delta_t$.

For definiteness, we perform all calculations for $M_{H^\pm}
=800$~GeV. In accordance with Eq.~(\ref{b2d2}), at
$M_\phi< 250$~GeV the contribution of the charged Higgs boson
loop varies by less than 5\% when $M_{H^{\pm}}$ varies from
800~GeV to infinity (for $M_{H^{\pm}}$ below 3~TeV, this is
well within the perturbative region).

The expected precision in the measurement of $\chi_{Z\gamma}$ is
lower than that for $\chi_{\gamma\gamma}$ as was noted above.
Besides, we find that the difference of the
$Z\gamma$ Higgs width from its SM value is lower than that for the
$\gamma\gamma$ width in all considered cases, i.e., the
$\phi Z\gamma$ coupling is less suitable for distinguishing the
discussed models than the $\phi\gamma\gamma$ coupling. Therefore,
the $\phi Z\gamma$ coupling measurements can only be a supplement
to those of $\phi\gamma\gamma$ for the considered problem.

The best place for measuring the $\phi Z\gamma$ coupling appears 
to be in the reaction $e\gamma\to e\phi$ \cite{GIv1}. 
Here the measurable quantity will be the interference 
${\rm Re} (\chi_{Z\gamma}^*\chi_{\gamma\gamma})$,
not for real $Z$ but for a $Z$ whose four-momentum is space-like. 
More detailed studies are needed.

%%%%%%%%%%%%%%%%%%%%%%%%%%%%%%%%%%%%%%%%%%%%%%%%%%%%%%%%%%%%%%%%%%%%%%%%
\subsection{Solutions A}\label{secA}
%%%%%%%%%%%%%%%%%%%%%%%%%%%%%%%%%%%%%%%%%%%%%%%%%%%%%%%%%%%%%%%%%%%%%%%%

%%%%%%%%%%%%%%%%%%%%%%%%%%%%%%%%%%%%%%%%%%%%%%%%%%%%%%%%%%%%%%%%%%%%%%%%
\subsubsection{The \boldmath$\gamma\gamma$ and $Z\gamma$ widths}
%%%%%%%%%%%%%%%%%%%%%%%%%%%%%%%%%%%%%%%%%%%%%%%%%%%%%%%%%%%%%%%%%%%%%%%%

A new feature of the $\gamma\gamma$ and $Z\gamma$ widths, 
as compared to the SM
case, is the contribution due to the charged Higgs boson loops.
It is known that the scalar loop contribution to the photonic
widths is less than that of the fermion and $W$ boson loops (the
last is the largest).  The contributions of $W$ and $t$-quark
loops are of opposite signs (see Eq.~(\ref{asym})), i.e.,
they partially compensate each other. Thus, the effect of scalar
loops is enhanced here.

According to Eq.~(\ref{b2d2}), the coupling $\chi_{H^{\pm}}$
depends linearly on $\lambda_5$. The variation of this coupling 
with $\lambda_5$ is small as long as 
$|\lambda_5|/ \lambda_4 \le {\cal O}(1)$.
For this case we found that one can write with high accuracy
\begin{equation}
|\chi_a|^2 =\frac{\Gamma_a^{\rm 2HDM}}{\Gamma_a^{\rm SM}}
=1-R_a\left(1-\frac{|\lambda_5|}{2\lambda_4}\right), \qquad 
a=\gamma\gamma\mbox{ or } Z\gamma, 
\label{Ra}
\end{equation}
with quantities $R_a$ which can be determined from $|\chi_a|^2$
at $\lambda_5=0$.

For $\lambda_5=0$ and $M_{H^\pm}= 800$ GeV the ratios of the
considered Higgs widths to their SM values are shown in
Fig.~\ref{Fig:sol-A}. The solid curves, which correspond to strict
SM values for the basic couplings, are below 1 due to the
contribution of the charged Higgs boson. The estimate for
$|\chi_{\gamma\gamma}|^2$ according to Eq.~(\ref{loop}) and the
asymptotic values (\ref{asym}) give for the solid curve
$R_{\gamma\gamma}=1-|\chi_{\gamma\gamma}|^2 = 1- (44/47)^2
\approx 0.12$. The precise calculation at $M_\phi=110$ GeV results
in the value 0.106, which is close to the asymptotic 
estimate.\footnote{A similar estimate for $Z\gamma$ width results
in $R_{Z\gamma}\approx 0.045$ which is close to the precise value 
0.041. This lower value of the deviation from unity for the 
$Z\gamma$ width is caused by the larger asymptotic $W$-loop 
contribution as compared to the asymptotic value for 
$a=\gamma\gamma$ [see Eq.~(\ref{asym})].}
The $R_a$ dependencies on $M_\phi$ are similar
for $a=\gamma\gamma$ and $a=Z\gamma$. 
For higher $M_\phi$ these
quantities decrease since the $W$-loop contribution $F_1(W)$
increases while the $t$-quark and $H^\pm$ contributions change 
more weakly.
The effect of the $W^+W^-$ threshold is clearly seen in the figures. 
At $M_\phi=250$ GeV we have $R_{\gamma\gamma}=0.05$ and 
$R_{Z\gamma}=0.018$.

%%%%%%%%%%%%%%%%%%%%%%%%%%%%%%%%%%%%%%%%%%%%%%%%%%%%%%%%%%%%%%%%%%%%%%%%
\begin{figure}[htb]
\refstepcounter{figure}
\label{Fig:sol-A}
\epsfig{file=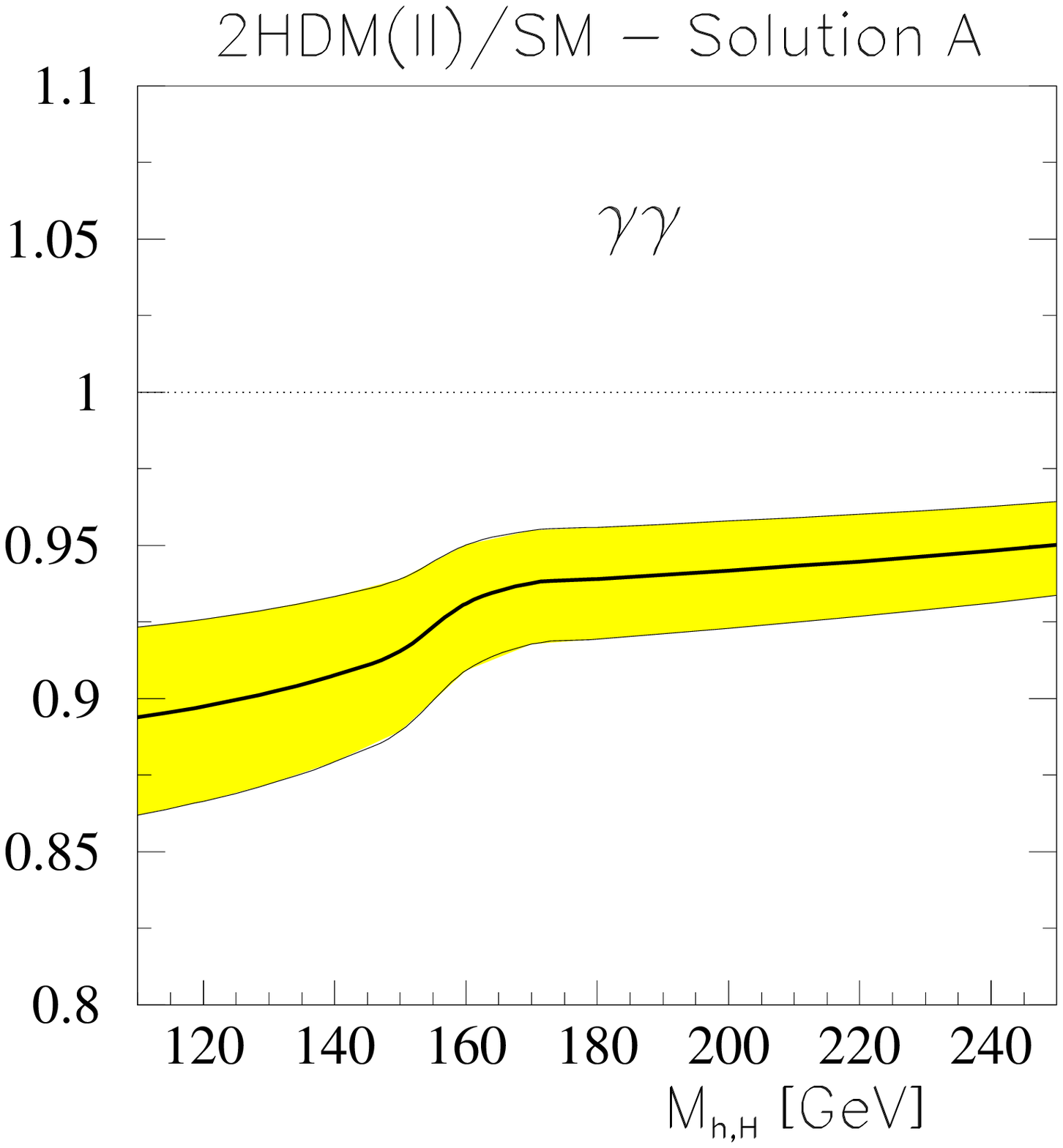,width=75mm}
\epsfig{file=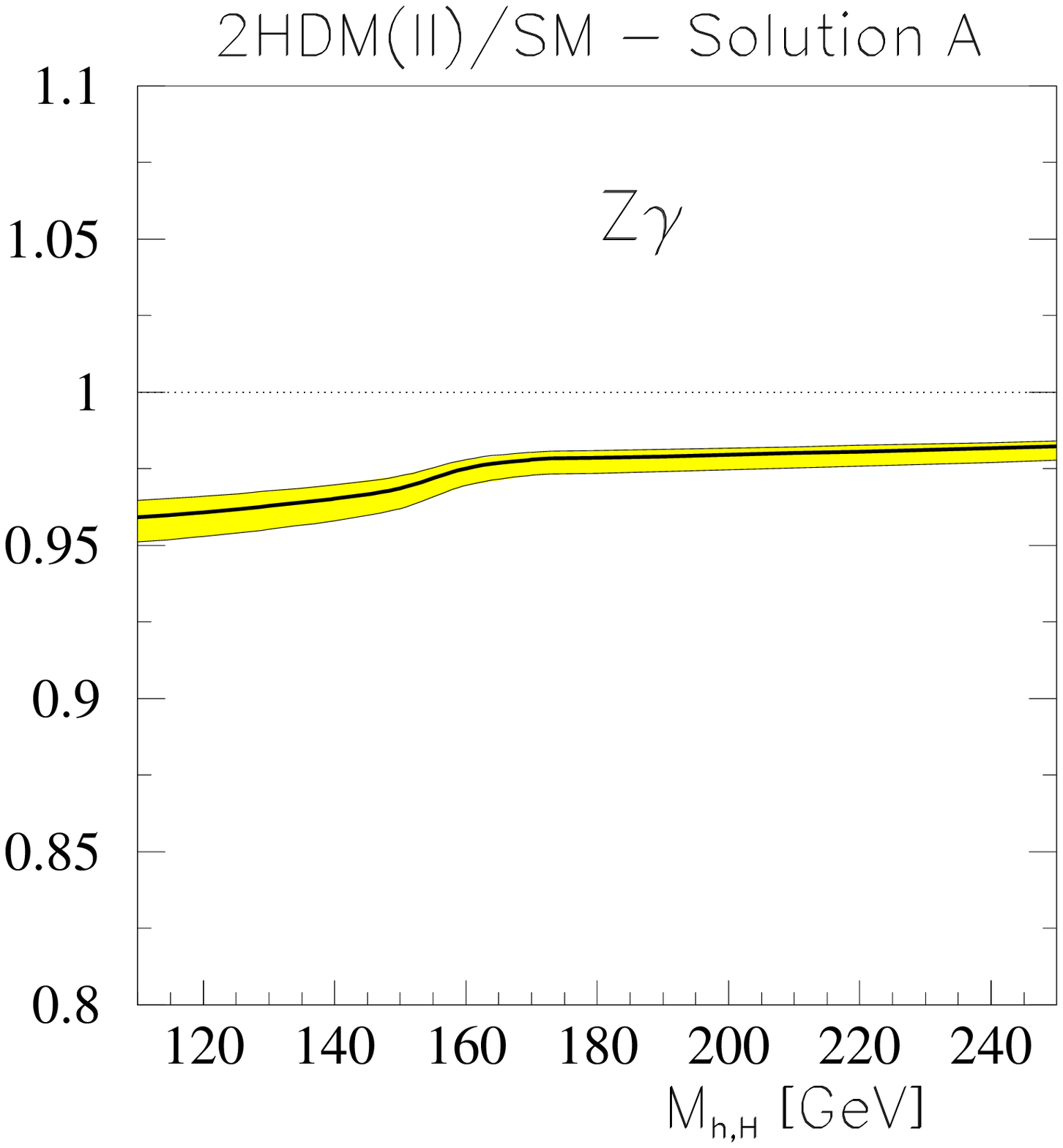,width=75mm}
\addtocounter{figure}{-1}
\caption{Ratios of the Higgs boson $\phi\to\gamma\gamma$ (left
panel) and $\phi\to Z\gamma$ (right panel) decay widths in the
2HDM~(II) and the SM as functions of $M_\phi$ for all solutions~$A$ with
$\lambda_5=0$. See text for a description of the solid curve
and the shaded bands.}
\end{figure}
%%%%%%%%%%%%%%%%%%%%%%%%%%%%%%%%%%%%%%%%%%%%%%%%%%%%%%%%%%%%%%%%%%%%%%%%

Since for these solutions $|\chi_V|=1$ with high accuracy and 
since the $b$-quark contribution is negligible, the uncertainty in 
the considered loop couplings $|\chi_a|^2$ (shown as shaded 
bands) is only due to the uncertainty in $\chi_u$ (which enters
with the coefficient $k_\gamma\approx |[8F_{1/2}(t)/3]|/ 
[F_1(W)+4F_{1/2}(t)/3] \approx 32/47\approx 2/3$ at small 
$M_h$ in $R_{\gamma\gamma}$ and a smaller coefficient $k_Z$ for 
$R_{Z\gamma}$). In accordance with the previous discussion, these 
$k_{\gamma}, k_{Z}$ become smaller at higher $M_h$.  The
shaded region above the solid curve corresponds to $\epsilon_u>0$ and
that below the solid curve to $\epsilon_u<0$. In accordance with the
discussion in Sec.~\ref{solAdescr}, both these regions are
relevant to solutions $A_{h+}$ (or $A_{H-}$) and only the lower 
region is relevant to the solution $A_{h-}$.

%%%%%%%%%%%%%%%%%%%%%%%%%%%%%%%%%%%%%%%%%%%%%%%%%%%%%%%%%%%%%%%%%%%%%%%%
\subsubsection{The two-gluon width}
%%%%%%%%%%%%%%%%%%%%%%%%%%%%%%%%%%%%%%%%%%%%%%%%%%%%%%%%%%%%%%%%%%%%%%%%

The two-gluon width is determined by the 
contributions of the $t$ and $b$ quark loops. 
For not too high values of $\tan\beta$, 
the $t$-quark contribution dominates. So, the difference
$\chi_{gg}-1$ is determined by the difference $\chi_u-1$, and we 
have checked that with high accuracy $\chi_{gg}-1 \approx 
2(\chi_u-1)=2\epsilon_u$. Let us consider now different regions 
of $\tan\beta$.\\
{\em (i)} If $\tan\beta\ll 1$, then the deviation of the Higgs
boson coupling with the $t$-quark from its SM value can be
significant, reaching the bounds of the expected experimental
uncertainty (\ref{couplacc}).\\
{\em (ii)} If $\tan\beta>1$, then according to Eq.~(\ref{sol2s}),
the deviation of the coupling with the $t$-quark from its SM
value should be less than that for the $b$-quark, the latter
being constrained from above by the experimental uncertainty
(\ref{couplacc}). Therefore, in this case one expects
$|\chi_{gg}-1| <0.05$, which is lower than the expected precision
in the measurement of the two-gluon width.

%%%%%%%%%%%%%%%%%%%%%%%%%%%%%%%%%%%%%%%%%%%%%%%%%%%%%%%%%%%%%%%%%%%%%%%%
\subsection{Solutions~B}\label{secB}
%%%%%%%%%%%%%%%%%%%%%%%%%%%%%%%%%%%%%%%%%%%%%%%%%%%%%%%%%%%%%%%%%%%%%%%%

For solutions $B$ we have, by definition, $\chi_u+\chi_d
={\cal O}(\epsilon)$. So with high accuracy $\chi_{H^\pm} \approx
\chi_V$ (see Eq.~(\ref{b2d2})), and the final result is independent 
of $\lambda_5$. The results of calculations of photon
widths are shown in Fig.~\ref{Fig:sol-B}, where the lower curves
correspond to the solutions $B_{\phi\pm d}$ and the upper ones to
the solution $B_{h+u}$.

%%%%%%%%%%%%%%%%%%%%%%%%%%%%%%%%%%%%%%%%%%%%%%%%%%%%%%%%%%%%%%%%%%%%%%%%
\subsubsection{The \boldmath$\gamma \gamma$ and $Z \gamma$ widths}
%%%%%%%%%%%%%%%%%%%%%%%%%%%%%%%%%%%%%%%%%%%%%%%%%%%%%%%%%%%%%%%%%%%%%%%%

{\em (i)} For the solutions~$B_{\phi\pm d}$ the main source of
deviation from the SM prediction is due to the charged Higgs
contribution and the small change due to the opposite relative
sign of the $b$-quark coupling ($\chi_d\simeq-\chi_V$) as 
compared to that in the SM case,
see Fig.~\ref{Fig:sol-B}, lower curves.
Therefore, the lower solid curves in Fig.~\ref{Fig:sol-B}
(calculated for $\chi_V=-\chi_d=\chi_u=\pm 1$) are very close
to those for solution $A$ in Fig.~\ref{Fig:sol-A}.
Varying the Higgs boson mass $M_{\phi}$ from 110 to 250 GeV we
obtain $|\chi_{\gamma\gamma}|^2 =0.860$ -- 0.957 and
$|\chi_{Z\gamma}|^2=0.953$ -- 0.984 for points on the solid curve.
The difference from the results for solution $A$ at $\lambda_5=0$ 
is determined mainly by the small $b$-quark loop contribution. 
Naturally, it decreases for larger values of $M_\phi$.

For these solutions the deviation of the $\phi t\bar{t}$ coupling
from its SM value is negligible (see Eq.~(\ref{solBds})). Besides,
since the contribution of the $b$-quark loop is small compared to
that of the $t$-quark loop, the photon widths are also insensitive
to the experimental uncertainty $\delta_b$. The only essential
dependence on the experimental inaccuracy is that of $\chi_V$. The
photonic widths are $\propto (1-\epsilon_V)^2$ with
$\epsilon_V>0$, so the allowed values of these widths can only be
lower than those at $\chi_V=\pm 1$. The width of the 
shaded regions is given by the corresponding factor $1-2\delta_V$.

%%%%%%%%%%%%%%%%%%%%%%%%%%%%%%%%%%%%%%%%%%%%%%%%%%%%%%%%%%%%%%%%%%%%%%%%
\begin{figure}[htb]
\refstepcounter{figure}
\label{Fig:sol-B}
\addtocounter{figure}{-1}
\epsfig{file=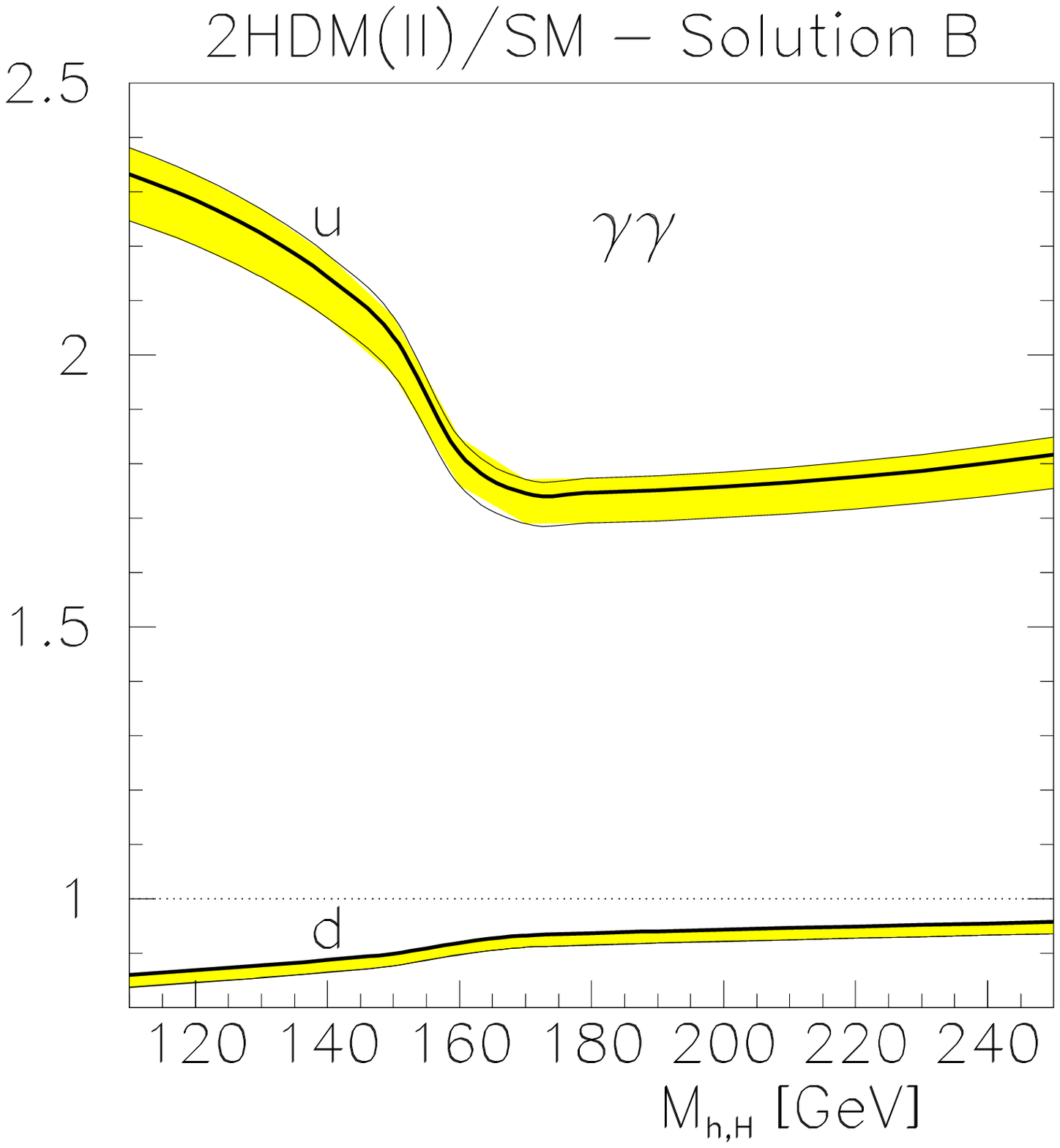,width=75mm}
\epsfig{file=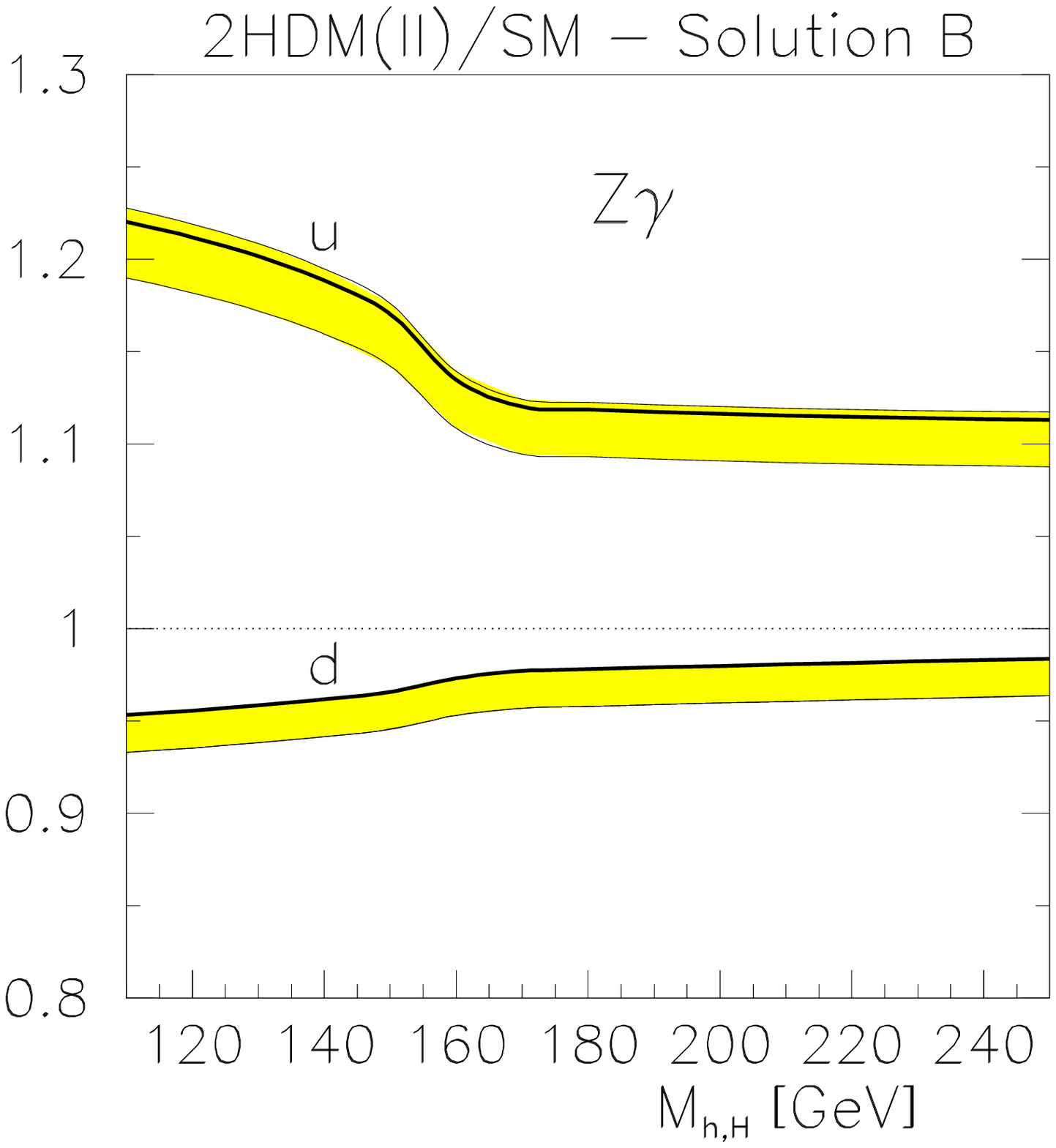,width=75mm} \caption{Ratios of the
Higgs boson $\phi\to\gamma\gamma$ (left panel) and $\phi\to
Z\gamma$ (right panel) decay widths in the 2HDM~(II) and the SM as
functions of $M_{\phi}$ for all solutions~$B$. The lower
bands correspond to solutions $B_{\phi\pm d}$. The upper
bands correspond to the solution $B_{h+u}$. See the text for a 
detailed description.}
\end{figure}
%%%%%%%%%%%%%%%%%%%%%%%%%%%%%%%%%%%%%%%%%%%%%%%%%%%%%%%%%%%%%%%%%%%%%%%%
{\em (ii)} For the solution $B_{h+u}$ the photon widths change
dramatically as compared to the SM case (upper parts
in Fig.~\ref{Fig:sol-B}). Here, solid curves correspond to the case
$\chi_V=\chi_d=-\chi_u= 1$.
With the asymptotic values
(\ref{loop}) and (\ref{asym}) we obtain $|\chi_{\gamma\gamma}|^2
\approx (76/47)^2\approx 2.6$. The small contribution of
the $b$-quark loop decreases this value. The numerical calculation
(upper solid curve in Fig.~\ref{Fig:sol-B}) gives here about 2.3 
for the ratio of the widths for $M_h=110$~GeV.
The change of behavior of this curve at $M_h\approx 2M_W$
corresponds to the change in the $M_h$ dependence of
the real part, and the onset of the imaginary part of the $W$-loop 
contribution.

For this solution the experimental uncertainty in the 
$ht\bar{t}$ coupling becomes essential. The upper shaded regions 
in the figure are derived numerically from the bounds around the 
SM values of two measured basic coupling constants,
$g_V$ ($V=Z$ or $W$) and $g_u$ ($=g_t$). The first uncertainty,
given by $\delta_Z$, allows for a reduction of 
$|\chi_{\gamma\gamma}|^2$. The second
uncertainty is reduced in the Higgs-two-photon coupling by a
factor $k_\gamma$ similar to that discussed for solution~$A$. In the
asymptotic region we have $k_\gamma\approx 32/76 \approx 0.4$.
This reduction factor $k_\gamma$ decreases with increasing values of
$M_h$. It gives the shaded regions, both below and above the solid
curve.

%%%%%%%%%%%%%%%%%%%%%%%%%%%%%%%%%%%%%%%%%%%%%%%%%%%%%%%%%%%%%%%%%%%%%%%%
\subsubsection{The two-gluon width}
%%%%%%%%%%%%%%%%%%%%%%%%%%%%%%%%%%%%%%%%%%%%%%%%%%%%%%%%%%%%%%%%%%%%%%%%

The two-gluon width has only quark loop contributions.
The results of numerical analyses are presented in
Fig.~\ref{Fig:sol-B-gg}. The solid curves correspond to the case
$\chi_u=-\chi_d=\pm 1$ for all solutions $B$. As in the
asymptotic cases (\ref{asym}), the loop contributions of $t$ and
$b$ quarks are of opposite signs and $|F_{1/2}(b)|<
|F_{1/2}(t)|$. In the SM case they partly cancel each other,
while for solutions $B$ these contributions add.
Therefore, for these solutions, the two-gluon width is
considerably higher than in the SM. With the asymptotic
Eqs.~(\ref{asym}) and setting $M_b(M_\phi)=4$ GeV we obtain for
$M_\phi=110$~GeV the value $|\chi_{gg}|^2\approx 1.4$ which is near the value
obtained by numerical calculation. Varying the Higgs mass from 110
to 250 GeV we obtain $|\chi_{gg}|^2= 1.31$ -- 1.09.

%%%%%%%%%%%%%%%%%%%%%%%%%%%%%%%%%%%%%%%%%%%%%%%%%%%%%%%%%%%%%%%%%%%%%%%%
\begin{figure}[htb]
\refstepcounter{figure} \label{Fig:sol-B-gg}
\addtocounter{figure}{-1}
\begin{center}
\epsfig{file=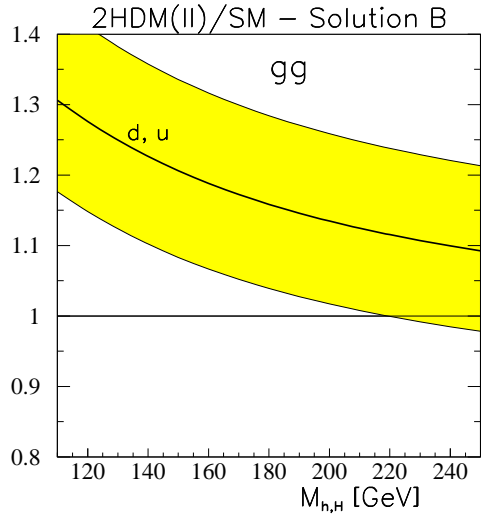,width=75mm} \caption{Ratio of the Higgs
boson $\phi\to gg$ decay widths in the 2HDM~(II) and the SM as
function of $M_{\phi}$ for all solutions $B_{\phi\pm d}$ and
$B_{h+u}$. The shaded bands correspond to the solution $B_{h+u}$
only (bands above or below the solid curve correspond to
$\epsilon_u<0$ or $\epsilon_u>0$).
For the solutions $B_{\phi\pm d}$, the shaded bands are practically
reduced to the width of the solid curve.}
\end{center}
\end{figure}
%%%%%%%%%%%%%%%%%%%%%%%%%%%%%%%%%%%%%%%%%%%%%%%%%%%%%%%%%%%%%%%%%%%%%%%%

According to Eq.~(\ref{solBds}), for the solutions $B_{\phi\pm d}$
the uncertainty in the $\phi t\bar{t}$ coupling is negligible while
the uncertainty in the $\phi b\bar{b}$ coupling is suppressed in
$\chi_{gg}$ since the $b$-quark loop contribution is small.
Therefore, for these solutions $B_{\phi\pm d}$, the shaded bands
in the figure are practically degenerated to zero width. 
However, for the solution $B_{h+u}$, the uncertainty induced
by the anticipated 1-$\sigma$ accuracy in the $ht\bar{t}$ coupling
is considerable, as indicated by the shaded bands, below and above
the solid curve.

%%%%%%%%%%%%%%%%%%%%%%%%%%%%%%%%%%%%%%%%%%%%%%%%%%%%%%%%%%%%%%%%%%%
\subsection{ Distinguishing models via loop-induced couplings}
\label{secdist}
%%%%%%%%%%%%%%%%%%%%%%%%%%%%%%%%%%%%%%%%%%%%%%%%%%%%%%%%%%%%%%%%%%%%

Our results for loop-induced couplings show no visible
difference in the results between the cases when the observed
Higgs boson is the lighter one, $h$, or the heavier one, $H$.
Only the (difficult) direct observation of $h$ could in the latter
case can rule out this possibility. Taking into account anticipated
experimental inaccuracies in the basic couplings of the Higgs
boson to fermions, to $Z$ and $W$ bosons and the freedom in the
values of $\lambda_5$, the discussed experiments cannot
discriminate between the solutions $A$ and $B$ either.

The different allowed realizations of an SM-like scenario 
discussed above, are summarized in Table~3.
Note that solutions $A_{h+}$ and $A_{H-}$ are similar in some respects,
as are $B_{h+d}$ and $B_{H-d}$.
In fact, they lead to the same predictions for the loop-induced
couplings, given in Figs.~1, 2 and 3.
Each of these solutions corresponds to two regions of $\epsilon_u$ 
($\epsilon_d$), except $A_{h-}$, where only the region
$\epsilon_u<0$ ($\epsilon_d>0$) is allowed.
In all these solutions the deviation $\epsilon_d$ can be either
positive or negative, except for solution $A_{h-}$ for which
$\epsilon_d$ should be negative.
%================================================================
\begin{table}[ht]
\begin{center}
\begin{minipage}{160mm}
\begin{tabular}{||c|c|c|c|c|c|c|c|c||}
\hline\hline
& $\chi_V=g_V/g_V^{\rm SM}$ 
& $\chi_d=g_d/g_d^{\rm SM}$ 
& $\chi_u=g_u/g_u^{\rm SM}$ & $\tan\beta$
& $|\chi_{\gamma\gamma}|^2$ & $|\chi_{Z\gamma}|^2$ 
& $|\chi_{gg}|^2$ \\
\hline\hline
$A_{h+}$ 
& $+(1-\epsilon_V)$ & $+(1-\epsilon_d)$ & $+(1-\epsilon_u)$ 
& $<1$, $>1$
& 0.90 & 0.96 & 1.00 \\
\hline
$A_{h-}$ 
& $-(1-\epsilon_V)$ & $-(1-\epsilon_d)$ & $-(1-\epsilon_u)$ 
& $\gg 1$
& 0.90 & 0.96 & 1.00 \\
\hline 
$A_{H-}$ 
& $-(1-\epsilon_V)$ & $-(1-\epsilon_d)$ & $-(1-\epsilon_u)$ 
& $>1$
& 0.90 & 0.96 & 1.00 \\
\hline\hline
$B_{h+d}$ 
& $+(1-\epsilon_V)$ & $-(1-\epsilon_d)$ & $+(1-\epsilon_u)$ 
& $>1$
& 0.87 & 0.96 & 1.28 \\
\hline
$B_{h-d}$ 
& $-(1-\epsilon_V)$ &  $+(1-\epsilon_d)$ & $-(1-\epsilon_u)$ 
& $>1$
& 0.87 & 0.96 & 1.28 \\
\hline
$B_{H-d}$ 
& $-(1-\epsilon_V)$ & $+(1-\epsilon_d)$ & $-(1-\epsilon_u)$ 
& $>1$
& 0.87 & 0.96 & 1.28 \\
\hline
$B_{h+u}$ 
& $+(1-\epsilon_V)$ & $+(1-\epsilon_d)$ & $-(1-\epsilon_u)$ 
& $<1$
& 2.28 & 1.21 & 1.28 \\
\hline\hline
\end{tabular}
\end{minipage}
\end{center}
\caption{SM-like scenarios in the 2HDM~(II).
The solutions correspond to $\epsilon_V>0$ and, for each realization, 
$\epsilon_d\epsilon_u<0$.
The ratios of partial widths (last three columns), refer
to the ``exact'' case ($\epsilon_i=0$), for $M_\phi=120$~GeV, 
and $\lambda_5=0$.}
\end{table}
%================================================================

%%%%%%%%%%%%%%%%%%%%%%%%%%%%%%%%%%%%%%%%%%%%%%%%%%%%%%%%%%%%%%%%%%%
\subsubsection{Distinguishing models at a Photon Collider}
%%%%%%%%%%%%%%%%%%%%%%%%%%%%%%%%%%%%%%%%%%%%%%%%%%%%%%%%%%%%%%%%%%%%

Our detailed estimates for loop-induced widths are related to 
the era after precise
measurements of Higgs boson coupling constants at an $e^+e^-$
Linear Collider with the uncertainties given by
Eq.~(\ref{couplacc}).

The comparison of the presented results for the two-photon width
with the anticipated experimental uncertainty (\ref{jik}) shows
that the deviation of the two-photon width ratio from unity
is generally large enough to allow a reliable distinction of
the 2HDM~(II) from the SM at a Photon Collider. 
For the solutions $B$ this
conclusion is valid for arbitrary values of $\lambda_5$ and
arbitrary masses of other (unobserved) Higgs bosons.
For the solutions $A$, according to Eq.~(\ref{Ra}), this
conclusion is valid in a wide range of $\lambda_5$ values,
except for the interval (in units of $2\lambda_4$)
$\sim(1-\delta_\gamma /R_{\gamma\gamma},
1+\delta_\gamma/R_{\gamma\gamma})$. The possible precision in the
determination of $\lambda_5v^2/M^2_{H^\pm}$ from the two-photon
width depends strongly on the achieved precision in the
determination of the $ht\bar{t}$ coupling. These conclusions
are accurate for a Higgs boson lighter than 140~GeV.
A new analysis of experimental uncertainties is necessary if
the observed Higgs boson will be heavier than 140~GeV.

The measurement of the $hZ\gamma$ coupling in the process
$e\gamma\to eh$ will be an additional test for the considered
problems.

%%%%%%%%%%%%%%%%%%%%%%%%%%%%%%%%%%%%%%%%%%%%%%%%%%%%%%%%%%%%%%%%%%%
\subsubsection{Distinguishing models at the Tevatron, LHC
and an $e^+e^-$ Linear Collider}
%%%%%%%%%%%%%%%%%%%%%%%%%%%%%%%%%%%%%%%%%%%%%%%%%%%%%%%%%%%%%%%%%%%%

Before a Photon Collider is realized, the SM-like scenario
can be considered with respect to data obtained first at the
Tevatron, then at the LHC, and finally at an $e^+e^-$ Linear 
Collider. The observation of loop-induced couplings can 
distinguish models in the frame of the ``current SM-like scenario'' 
determined via currently measured coupling constants. 
Even at the Tevatron the solution $B_{h+u}$ can easily be 
distinguished via a study of the process $gg \to \phi \to
\gamma\gamma$ with rate about three times higher than
that in the SM (the product of the ratios of decay widths
presented in Figs.~3 and 2 (left panel, upper curve)).
Such an enhancement is estimated (in some other respect) 
as observable at the Tevatron \cite{ggggam}.

For all other solutions the possibility of distinguishing
models before the Photon Collider era is also related to the study of
the two-gluon width of the Higgs boson. Unfortunately, the
expected experimental precision in the measurement of this
quantity is not so high, and the ambiguity in its calculation is
also essential. Therefore, the results obtained via the two-gluon
width should be supplemented by some independent experiment,
such as the two-photon width measured via photon fusion at
a Photon Collider.

For all solutions $B_{\phi\pm}$ the two-gluon width is about 30\%
higher than its SM value at $M_\phi=115$ GeV and decreases fast
for increasing values of $M_\phi$ (due to the change of relative
sign of the $t$ and $b$ loop contributions as compared to the SM
case). If such a precision can be achieved, it helps to
distinguish models via the two-gluon width.

For solutions $A$ the deviation of the two-gluon width in the
2HDM~(II) from its SM value is twice as large as the deviation in
the Higgs boson coupling to $t$ quarks. The last deviation can be
high if  $\tan\beta\ll 1$. In this case it can happen that the
deviation of the two-gluon width from its SM value will be observable
while the deviation in the coupling to $t$ quarks will lie below
the experimental resolution. Note that the possible decrease of
the two-gluon width as compared to the SM value can be realized only
in solutions $A$. (The value $|\chi_{gg}|^2>1$ can be realized in
both solutions $A$ and $B$.)

%%%%%%%%%%%%%%%%%%%%%%%%%%%%%%%%%%%%%%%%%%%%%%%%%%%%%%%%%%%%%%%%%%%%%%%%
\subsection{Distinguishing the MSSM from the SM}
%%%%%%%%%%%%%%%%%%%%%%%%%%%%%%%%%%%%%%%%%%%%%%%%%%%%%%%%%%%%%%%%%%%%%%%%

In a forthcoming publication we plan to study the MSSM in
this same SM-like scenario. In this model only some of the
solutions $A$ of the pattern relation can be realized, and the
additional constraint on $\lambda_5$ ($\approx 2\lambda_4$) makes
the contribution of the charged Higgs boson to the discussed
photon width very small. On the other hand, the contributions 
of superpartners can be substantial.  
Since the mechanism of generation of their
masses differs from that in the SM, superparticles are completely
decoupled from the Higgs boson if they are sufficiently heavy. 

The question whether or not it is possible to see signals of the MSSM
as compared to the SM in this decoupling limit, was considered 
in Ref.~\cite{Djouadi1}.
The two-photon width was studied for finite values of
superparticle masses, but using an old estimate for the uncertainty 
in BR$(h\to \gamma\gamma)$, 10\%,
instead of the modern value (\ref{jik}). With that uncertainty, it
was concluded that the MSSM and the SM cannot be resolved via the
two-photon width even for light chargino and top squark (250~GeV).
These estimates should be reconsidered to obtain real bounds,
relevant to the uncertainty (\ref{jik}).
Also, one should consider masses of superpartners and other 
parameters of the theory at which the models cannot be distinguished 
in other ways. 
Such values could be above the discovery limits for the LHC.

%%%%%%%%%%%%%%%%%%%%%%%%%%%%%%%%%%%%%%%%%%%%%%%%%%%%%%%%%%%%%%%%%%%%%%%%
\section{Conclusion}
%%%%%%%%%%%%%%%%%%%%%%%%%%%%%%%%%%%%%%%%%%%%%%%%%%%%%%%%%%%%%%%%%%%%%%%%

We consider the case that after the Higgs boson discovery no
signal of New Physics will be found (SM-like scenario). 
This can occur both in the SM and in other models,
including the 2HDM~(II), MSSM, etc. 
Realization of an SM-like scenario constrains these alternative
models strongly, and measurements of loop-induced couplings
of the observed Higgs boson can help to understand which model 
is realized.

In this paper we consider this problem in the 2HDM~(II) for a
Higgs boson (which can be either one of two neutral scalars,
$h$ or $H$). For this purpose, we consider
the ratios of measurable couplings of the Higgs boson with
quarks and electroweak gauge bosons to their SM values. We
obtain constraints on the 2HDM~(II) in the form of a
pattern relation among these ratios. (These particular
pattern relations are valid also in the MSSM.)

We analyzed two possible classes of solutions of the pattern
relation for the considered SM-like scenario. In the solutions
$A$ all basic couplings are close to their SM values. In the
solutions $B$ some of the basic couplings are close to their SM
values while others differ in sign from the SM values. Next, we
calculated the $\gamma\gamma$ and $Z\gamma$ partial widths of
the observed Higgs boson for all considered solutions, taking
into account anticipated uncertainties in future measurements of
the basic couplings of the Higgs boson.

The results of our calculations are shown in the figures. 
The solid curves describe an ``exact'' SM-like scenario with basic
couplings $\chi_i=\pm 1$. The shaded bands represent
effects of experimental uncertainties in the measurement of
basic couplings at an $e^+e^-$ Linear Collider for $M_\phi<140$~GeV
(\ref{couplacc}).

If $M_\phi>140$ GeV, these estimates of shaded bands should be
reconsidered. In this region the dominant decay channels of SM-like
Higgs boson changes (becoming $WW^*$ and $ZZ^*$) just as the main
mechanism of its production at an $e^+e^-$ collider changes ($W$ fusion
instead of Higgs-strahlung). Therefore, the values of the
uncertainties of couplings of the Higgs boson to quarks, $W$ and $Z$
will be changed.

Our main results are related to the era after the LHC and
$e^+e^-$ Linear Collider operations if the SM-like scenario 
will be found to be realized at that time. 
We found the difference between the SM and
the 2HDM~(II) to be large enough to discriminate the models via
measurements of Higgs boson production at a Photon Collider ---
for solutions $B$ in general, and for solutions $A$ except a 
limited range of values of $\lambda_5$.

For the period of operations at the Tevatron, LHC and an
$e^+e^-$ Linear Collider we obtain the regions in the parameter
space where the basic couplings squared are close to the SM
case but the loop-induced
two-gluon or two-photon widths differ so significantly from
their SM values that one can distinguish the 2HDM~(II) from the
SM via experiments at the LHC or at an $e^+e^-$ Linear Collider.

\medskip

We are grateful to A. Djouadi, J. Gunion, H. Haber and M. Spira
for discussions of decoupling in the 2HDM and the MSSM,
and to P.~Chankowski,
W.~Hollik and I.~Ivanov for valuable discussions of the parameters
of the 2HDM. MK is grateful to the University of Bergen and to
the Theory Group at DESY for warm hospitality and financial support.
This research has been supported by RFBR grants 99-02-17211 and
00-15-96691, by Polish Committee for Scientific Research, grant
No.\ 2P03B01414, and by the Research Council of Norway.

%\clearpage

\end{document}